\documentclass{ws-ijgmmp}
\usepackage[T1]{fontenc}
\usepackage[latin1]{inputenc}
\usepackage{graphicx}
\usepackage[english]{babel}
\usepackage{graphicx}
\usepackage{bm}
\usepackage{amsfonts}
\usepackage{amsmath}
\usepackage{amssymb}
\usepackage{graphicx}
\usepackage{epsfig}
\usepackage{colordvi}
\usepackage{color}

\providecommand{\U}[1]{\protect\rule{.1in}{.1in}}
\begin{document}

\title{ THE AFFINE STRUCTURE OF GRAVITATIONAL THEORIES: SYMPLECTIC GROUPS AND
GEOMETRY}
\author{S.  Capozziello$^{1,2}$,  D. J. Cirilo-Lombardo$^{3}$, 
M.  De Laurentis$^{1,2}$}

\address{$^1$Dipartimento di Fisica, Università
di Napoli,  {}``Federico II'', Compl. Univ. di
Monte S. Angelo, Edificio G, Via Cinthia, I-80126, Napoli, Italy,\\
 $^2$INFN Sez. di Napoli, Compl. Univ. di
Monte S. Angelo, Edificio G,\\ Via Cinthia, I-80126, Napoli, Italy,\\
$^3$Bogoliubov Laboratory of Theoretical Physics, Joint Institute for
Nuclear Research,\\
141980, Dubna (Moscow Region), Russian Federation.}

\maketitle

\begin{abstract}
We give a geometrical description of  gravitational
theories from the  viewpoint of symmetries and affine structure. We show
how  gravity,  considered as a gauge theory,  can be
consistently achieved by the nonlinear realization of the conformal-affine
group in an indirect manner: due the partial isomorphism between $CA\left(
3,1\right) $ and the centrally extended $Sp\left( 8\right) $,  we perform a
nonlinear realization of the centrally extended (CE)$Sp\left( 8\right) $ in
its semi-simple version. In particular, starting from the bundle structure of gravity, we derive the conformal-affine Lie algebra and then, by the non-linear realization, we define the coset field transformations, the Cartan forms and the inverse Higgs constrants. Finally we discuss the geometrical Lagrangians where all the information on matter fields and their interactions can be contained.
\end{abstract}


\markboth{S.  Capozziello, D. J. Cirilo-Lombardo, M. De
Laurentis}
{The affine structure of  gravitational theories}

%
\catchline{}{}{}{}{} %

\begin{history}
\received{(Day Month Year)}
\revised{(Day Month Year)}
\end{history}

\section{Introduction}

The  Standard Model (SM) of particles is a gauge theory, where all
 fields mediating  interactions are represented by gauge potentials.  A very debated
question is to understand why the fields mediating the gravitational
interaction are different from those of  other fundamental forces. It is
reasonable to expect that there may be a gauge theory where the
gravitational fields stand on the same footing as those of the other fundamental fields.
This expectation has prompted a re-examination of General Relativity (GR) from the gauge field
viewpoint.

In the SM, the involved gauge groups  are internal symmetries
while, in GR, gauge groups  are associated to external spacetime symmetries. Since  GR
is not  a true Yang-Mills theory, we must require that it is a theory where gauge objects are not only the gauge potentials but also 
tetrads that relate the symmetry group to the external spacetime. For this reason, we
have to take into account  more complex nonlinear gauge theories. 

Ashtekar \cite{ashtekar} considered as fundamental variables the tetrad fields and the
connection forms while Einstein took,  as the fundamental variables, the spacetime metric components.  However
 gravitation may be viewed as a gauge theory \cite{uty} in analogy to
the Yang-Mills theory \cite{yang}. Later, a gauge theory build with Poincar\'e
group $P(3,1)=T(3,1)\rtimes SO(3,1)$ was introduced by Kibble
\footnote{Here $\rtimes$ represents the semi-direct product.} \cite{kibble}. Cartan 
\cite{cartan} generalized the Riemann geometry to include torsion in
addition to curvature. Many authors stressed that intrinsic spin may be the
source of torsion of the underlying spacetime manifold \cite{Sciama,Finkelstein,Hehl,Hehl1} and  gauge theories based on Lie groups have
been widely considered \cite{Hehl,Hehl1,Mansouri,Mansouri,Grignani}. 

Furthermore,
 it has been proposed that the Kibble gauge theory can be
built considering the de Sitter group $SO(4,1)$, \textit{i.e.} one has to come into the
Poincar\'e group by a group contraction  \cite{Inomata}. 

Usually one adopts the fiber-bundle
description to derive gauge theories on the Lie group, where the tetrads are
constructed by using the affine group $A(4,\mathbb{R})=T(4)\rtimes GL(4,\mathbb{%
R})$. It is important to stress that, in metric-affine gravity, the Lagrangian can be assumed quadratic in both
curvature and torsion terms in contrast to the Einstein-Hilbert Lagrangian of GR
which is linear in the scalar curvature. This approach has been recently
developed also for more general theories like $f(R)$-gravity \cite{Capozziello,Cianci, Magnano,PRnostro,faraoni,review1,review5}. However, there are many attempts to formulate gravitation as a gauge
theory.  Despite  of these  approaches, no final theory can be  uniquely accepted as the gauge theory
of gravitation. 

Besides,  the
non-linear approach in the context of internal symmetry groups has been considered \cite{Callan,Coleman} and extended to the case of spacetime symmetries, using the
nonlinear action of $GL(4,\mathbb{R})$ \cite{Isham,Salam,Salam1}. Other authors  \cite{Borisov,Ivanov} considered the simultaneous
nonlinear realization of the affine and conformal groups, showing that GR
can be viewed as a consequence of spontaneous symmetry breaking of the affine
symmetry where the gravitons are considered as Goldstone bosons associated
with the affine symmetry breaking. The nonlinear realization scheme,
employing $GL(4,\mathbb{R})$ as the principal group,  has been also considered  in \cite{Chang}, while the nonlinear realization induced by the
spontaneous breakdown of $SO(3,2)$ was investigated by Stelle et al. \cite{Stelle}. 

After, nonlinear gauge theories of the Poincar\'e, de Sitter,
conformal and special conformal groups have been taken into account \cite{Ivanov2,Ivanov3}
with the gravity as a spontaneously broken $GL(4,\mathbb{R})$ gauge theory.

Finally, the nonlinear realization in the fiber bundle formalism, based on
the bundle structure $G(G/H,H)$  has been  developed by several authors  (see for example \cite{Regge,Lord,Lord1}). In this
approach,  the quotient space $G/H$ is identified with physical spacetime.
Most recently,  nonlinear gauge theories of gravity have been discussed
on the basis of the Poincar\'e, affine and conformal groups \cite{37,38,39,40,42,43}. While in GR the spacetime is represented by  a
four-dimensional differential manifold that is assumed to be curved, in Special
Relativity (SR) the manifold is represented by the flat-spacetime $M_{4}$ (%
\textit{i.e.} Minkowskian spacetime). The GR can be regarded as a gauge
theory which is based on the local Lorentz group as well as, the
Yang-Mills gauge theory is based on the internal iso-spin gauge group.  From this viewpoint,
the Riemannian connection is the gravitational counterpart of the Yang-
Mills gauge fields. While $SU(2)$, in the Yang-Mills theory, is an internal
symmetry group, the Lorentz symmetry represents the local nature of
spacetime rather than internal degrees of freedom. 

On the other hand, the Einstein Equivalence
Principle, asserted for GR, requires that the local spacetime structure can
be identified with the Minkowski spacetime possessing the  Lorentz symmetry. In
order to connect the external spacetime to the local Lorentz symmetry, we need
to tie the external space to the local space. The tetrad fields can be the
link tools. The tetrads can be  considered as  objects that can be dynamically
generated and the spacetime must contain torsion in order to have spinor
fields. Namely, the gravitational interaction of spinning particles needs
that the Riemann spacetime must be modified in a (non-Riemannian) curved
spacetime with torsion. The Sciama theory fails when the tetrad fields are
treated as  gauge fields.  However, starting from a purely gauge point of
view, it is possible formulate a self-consistent 
gravitational theory   \cite{kibble,reviewFOOP}.

In this paper we discuss  how  gravity can be obtained as
a gauge theory achieved by the nonlinear realization of the
conformal-affine group in  an indirect manner: due to the partial isomorphism
between $CA\left( 3,1\right) $ and $Sp\left( 8\right)$,  we perform a
nonlinear realization of the centrally extended (CE)$Sp\left( 8\right)$in
its semisimple version. In order to obtain this  result, we consider the structure of the $CA(3,1)$, 
that is $SO(4,2)\cup A\left( 4,\mathbb{R}\right) $. It can be described
schematically as the 15 generators of the conformal group: Lorentz (6),
translations (4), special conformal (4) and dilatations (1), plus the 9
traceless generators of the symmetric linear transfomations (shear). Notice
that the copies of the common generators between $SO(4,2)$ and $A\left( 4,%
\mathbb{R}\right) $ have been eliminated.

The relation with the centrally extended $sp(8)$ algebra is based on the
observation that the nine traceless generators of the symmetric linear
transfomations (shear) can be factored in the eight parameter generators $%
F_{\rho\dot{\tau}}$, $\overline{F}_{\rho\dot{\tau}}\equiv(F_{\tau\dot{\rho}%
})^{\ast}\,$\ plus the axion $A$. This fact  brings us to the possibility to
make a nonlinear realization of the CE $Sp(8),$ where, due the partial
isomorphism described before,  conformal affine structure of the
gravitational theories can be included  \cite{reviewFOOP}.

A nonlinear realization is performed by the standard CE $sp(8)$ algebra:
the subgroup structure is  generated by $\widehat{K}_{\alpha\overset{.}{\alpha}}$
$\widehat{\widetilde{Z}}_{\alpha\beta},X_{\alpha\overset{.}{\alpha}}$,  where
the Lorentz generators $L_{\alpha\lambda }$ are  semisimple, in 
contrast with the subgroup structure generated by $K_{\alpha\overset{.}{\alpha}}$ 
$\widetilde{Z}_{\alpha\beta},F_{\alpha \overset{.}{\alpha}}$ $%
Z_{\alpha\beta},\overline{F}_{\alpha\overset{.}{\alpha }}$ where the Lorentz
generators $L_{\alpha\lambda}$  are   semidirect products. The main
difference between the different structures of \ $sp(8)$ consist in  the explicit
nonlinear constraints arising after the imposition that the corresponding
Cartan form of the scalar (pseudoscalar) field must be zero. This simple
fact makes a sharp difference between the simplest geometrical Lagrangians
(measures) obtained in each case: Yang-Mills type or Born-Infeld
(Nambu-Goto) type, that is: 

\bigskip%
\begin{equation*}
\begin{array}{ccc}
   &  & semisimple\rightarrow L_{G}:Born-Infeld/Nambu-Goto \\ 
sp(8) & 
\begin{array}{c}
\nearrow \\ 
\searrow%
\end{array}
&   \\ 
& &semidirect\text{ }product\rightarrow L_{G}:Yang-Mills%
\end{array}%
\end{equation*}
The layout of the paper is the following. 
Section \ref{tre} is
devoted to a general  discussion of the bundle structure of gravitation, while the
conformal-affine Lie algebra is introduced in Section \ref{quattro}. The
non-linear realization and the coset field transformations are discussed in
Section \ref{cinque}.  Relevant  Cartan forms are discussed  in
Section \ref{sei}, and the inverse Higgs
constraints are derived  Section \ref{sette} . The affine structure of superpsace is presented in Section \ref
{otto}. Finally, we shown the geometrical Lagrangian of the theory and give the new
symmetries and dynamics in Section \ref{nove}. Conclusions and perspectives 
are presented in Section \ref{dieci}.


\section{The Bundle Structure of Gravity}

\label{tre} 

Let us  briefly describe the bundle structure of gravity. Let $%
\mathbb{P}(M$, $G$; $\pi$) be a principal fiber bundle, where $M$ is the base space
and $G$  a standard diffeomorphic fiber. It  follows  that the gauge
transformations are characterized by the bundle isomorphisms \cite{Giachetta} $%
\lambda:\mathbb{P}\rightarrow\mathbb{P}$ exhausting all diffeomorphisms $%
\lambda_{M}$ on $M$. If a mapping is equivariant with respect to the action
of $G$, it is called an automorphism of $\mathbb{P}$. This amounts to restrict the action $\lambda$ of $G$ along fibers not altering the base space.
As it is well known, a gauge transformation is a fiber preserving
the bundle automorphisms, \textit{i.e.} diffeomorphisms $\lambda$ with $%
\lambda_{M}=\left( id\right) _{M}$. The automorphisms $\lambda$ form a group
called the automorphism group $Aut_{\mathbb{P}}$ of $\mathbb{P}$. The gauge
transformations form a subgroup of $Aut_{\mathbb{P}}$ called the gauge group 
$G\left( Aut_{\mathbb{P}}\right) $ (or $G$ in short) of $\mathbb{P}$. The
map $\lambda$ is required to satisfy two conditions, namely its
commutability with the right action of $G$ $[$the equivariance condition $%
\lambda\left( R_{g}(p)\right) =\lambda\left( pg\right) =\lambda\left(
p\right) g]$%
\begin{equation}
\lambda\circ R_{g}(p)=R_{g}(p)\circ\lambda\text{, \ }p\in\mathbb{P}\text{, }%
g\in G\,,
\end{equation}
according to which fibers are mapped into fibers, and the verticality
condition%
\begin{equation}
\pi\circ\lambda\left( u\right) =\pi\left( u\right) \text{,}
\end{equation}
where $u$ and $\lambda\left( u\right) $ belong to the same fiber. The last
condition ensures that no diffeomorphisms $\lambda_{M}:M\rightarrow M$, given
by%
\begin{equation}
\lambda_{M}\circ\pi\left( u\right) =\pi\circ\lambda\left( u\right) \text{,}
\end{equation}
is  allowed on the base space $M$.
To get a gauge description of gravity, we have to gauge external transformation groups. Hence, the transformations in a space must induce
corresponding transformations in the other. The usual definition of a gauge
transformation, \textit{i.e.} as a displacement along local fibers not
affecting the base space, must be generalized to reflect this interlocking.
One possible way of framing this interlocking is to employ a nonlinear
realization of the gauge group $G$, provided that a closed subgroup $H\subset G$
exists. The interlocking requirement is then transformed into the interplay
between groups $G$ and one of its closed subgroups $H$ \cite{reviewFOOP}.

Let us denote by $G$ a Lie group with elements $\left\{ g\right\} $. Let $H$ be a
closed subgroup of $G$ specified by 
\begin{equation}
H:=\left\{ h\in G|\Pi\left( R_{h}g\right) =\pi\left( g\right) \text{, }%
\forall g\in G\right\} \text{,}
\end{equation}
with elements $\left\{ h\right\} $ and known linear representations $%
\rho\left( h\right) $. Here $\Pi$ is the first of the two projection maps,
and $R_{h}$ is the right group action. Let $M$ be a differentiable manifold
with points $\left\{ x\right\} $ to which $G$ and $H$ may be referred, 
\textit{i.e.} $g=g(x)$ and $h=h(x)$. Being that $G$ and $H$ are Lie groups,
they are also manifolds. The right action of $H$ on $G$ induces a complete
partition of $G$ into mutually disjoint orbits $gH$. Since $g=g(x)$, all
elements of $gH=\left\{ gh_{1}\text{, }gh_{2}\text{, }gh_{3}\text{,}%
\cdot\cdot\cdot\text{, }gh_{n}\right\} $ are defined over the same $x$.
Thus, each orbit $gH$ constitutes an equivalence class of point $x$, with
equivalence relation $g\equiv g^{\prime}$, where\ $g^{\prime}=R_{h}g=gh$. By
projecting each equivalence class onto a single element of the quotient
space $\mathcal{M}:=G/H$, the group $G$ becomes organized as a fiber bundle
\ in\ the sense that $G=\bigcup\nolimits_{i}\left\{ g_{i}H\right\} $. In
this way,  the manifold $G$ is viewed as a fiber bundle $G\left( \mathcal{M}%
\text{, }H\text{; }\Pi\right) $ with $H$-diffeomorphic fibers $%
\Pi^{-1}\left( \xi\right) :G\rightarrow\mathcal{M}=gH$ and base space $%
\mathcal{M}$. A composite principal fiber bundle\textit{\ }$\mathbb{P}(M$, $%
G $; $\pi)$ is one whose $G$-diffeomorphic fibers possess the fibered
structure $G\left( \mathcal{M}\text{, }H\text{; }\Pi\right) \simeq\mathcal{M}%
\times$ $H$ described above. The bundle $\mathbb{P}$ is then locally
isomorphic to $M\times G\left( \mathcal{M}\text{, }H\right) $. Moreover,
since an element $g\in G$ is locally homeomorphic to $\mathcal{M}\times H$, 
the elements of $\mathbb{P}$ are - by transitivity - also locally
homeomorphic to $M\times\mathcal{M}\times H\simeq\Sigma\times H$ where
(locally) $\Sigma\simeq M\times\mathcal{M}$. Thus, an alternative view \cite%
{40} of $\mathbb{P}(M$, $G$; $\pi)$ is provided by the $\mathbb{P}$%
-associated $H$-bundle $\mathbb{P}(\Sigma$, $H$; $\widetilde{\pi})$. The
total space $\mathbb{P}$ may be regarded as $G\left( \mathcal{M}\text{, }H%
\text{; }\Pi\right) $-bundles over base space $M$ or equivalently as $H$%
-fibers attached to manifold $\Sigma\simeq M\times\mathcal{M}$.

The nonlinear realization (NLR) technique \cite{Callan, Coleman} gives us a
method to construct the transformation properties of fields defined on the
quotient space $G/H$. Thank to the  Ogievetsky theorem,  we can treat the
NLR of Diff$\left(4\text{, } \mathbb{R}\right)$, and then the algebras of Diff$%
\left(4\text{, } \mathbb{R}\right)$ can be considerate as algebras of $SO(4$%
, $2)$ and $A(4$, $\mathbb{R})$ \cite{Borisov}. We remember that the
trasformations with quadradic forms on the Minkowski spacetime are generated by the 
Lorentz group while infinitesimal angle-preserving transformations on the 
Minkowski spacetime are generated by special conformal groups. A
generalization of Poincar\'e group is the affine group and therefore the
Lorentz group is substituted by the group of general linear transformations 
\cite{felix}. As such, with the affine group, we can generate Lorentz
transformations, translations, volume preserving shear and volume changing
dilation transformations.

As a consequence, the NLR of Diff$\left( 4\text{, }%
\mathbb{R}
\right) /SO(3$, $1)$ can be constructed by taking a simultaneous realization
of the conformal group $SO(4$, $2)$ and the affine group $A(4$, $%
\mathbb{R}
):=%
\mathbb{R}
^{4}\rtimes GL(4$, $%
\mathbb{R}
)$ on the coset spaces $A(4$, $%
\mathbb{R}
)/SO(3$, $1)$ and $SO(4$, $2)/SO(3$, $1)$. Therefore, we can see that the
conformal-affine group could be the subgroup of Diff$\left( 4, \mathbb{R}
\right)$.  On the other hand, a group of realization is not linear when it is
subject to constraints. For example, we can choose constraints as those
responsible for the reduction of symmetry from Diff$\left( 4,\mathbb{R}
\right) $ to $SO(3$, $1)$. We consider the group $CA(3$, $1)$ as the basic
symmetry group $G$. The conformal-affine group consists of the groups $SO(4$%
, $2)$ and $A(4$, $%
\mathbb{R}
)$. In particular, the conformal-affine group is proportional to the union $SO(4$, $%
2)\cup A(4$, $%
\mathbb{R}
)$. We know, however, that the affine and special conformal groups have
several group generators in common. These common generators reside in the
intersection $SO(4$, $2)\cap A(4$, $%
\mathbb{R}
)$ of the two groups, within which there are \textit{two copies} of $%
\Pi:=D\times P(3$, $1)$, where $D$ is the group of scale transformations
(dilations)\ and $P(3$, $1):=T\left( 3\text{, }1\right) \rtimes SO(3$, $1)$
is the Poincar\'e group. We define the conformal-affine group as the union of
the affine and conformal groups minus \textit{one copy} of the overlap $\Pi$%
, i.e. $CA(3$, $1):=SO(4$, $2)\cup A(4$, $%
\mathbb{R}
)-\Pi$. Being defined in this way,  we recognize that $CA(3$, $1)$ is a $24$
parameter Lie group representing the action of Lorentz transformations $(6)$, translations $(4)$, special conformal transformations $(4)$, spacetime
shears $(9)$ and scale transformations $(1)$. Here, we have calculated the
NLR of $CA(3$, $1)$ modulo $SO(3$, $1)$.


\section{The Conformal-Affine Lie Algebra}

\label{quattro} 

In order to develop the NLR procedure, we choose to perform the partition of Diff$(4, 
\mathbb{R} )$ with respect to the Lorentz group. Using the Ogievetsky
theorem \cite{Borisov}, we identify representations of Diff$(4, \mathbb{R}
)/SO(3,1)$ with those of $CA(3$, $1)/SO(3$, $1)$. The $20$ generators of
affine transformations can be decomposed into the $4$ translational $\mathbf{%
P}_{\mu}^{\text{Aff}}$ and $16$ $GL(4, \mathbb{R})$ transformations $\mathbf{%
\Lambda}_{\alpha}^{\text{ }\beta}$. The $16$ generators $\mathbf{\Lambda}%
_{\alpha}^{\text{ }\beta}$ may be further decomposed into the $6$ Lorentz
generators $\mathbf{L}_{\alpha}^{\text{ }\beta}$ plus the remaining $10$
generators of symmetric linear transformation $\mathbf{S}_{\alpha}^{\text{ }%
\beta}$, that is, $\mathbf{\Lambda}_{\text{ }\beta}^{\alpha}=\mathbf{L}_{%
\text{ }\beta}^{\alpha}+\mathbf{S}_{\text{ }\beta}^{\alpha}$. The $10$
parameter symmetric linear generators $\mathbf{S}_{\alpha}^{\text{ }\beta}$
can be factored into the $9$ parameter shear (the traceless part of $\mathbf{%
S}_{\alpha}^{\text{ }\beta}$) generators defined by $^{\dagger}\mathbf{S}%
_{\alpha}^{\text{ }\beta}=\mathbf{S}_{\alpha }^{\text{ }\beta}-\frac{1}{4}%
\delta_{\alpha}^{\text{ }\beta}\mathbf{D}$, and the $1$ parameter dilaton
generator $\mathbf{D}=tr\left( \mathbf{S}_{\alpha }^{\text{ }\beta}\right) $%
. Shear transformations generated by $^{\dagger }\mathbf{S}_{\alpha}^{\text{ 
}\beta}$ describe shape changing, volume preserving deformations, while the
dilaton generator gives rise to volume changing transformations. The 4
diagonal elements of $\mathbf{S}_{\alpha }^{\text{ }\beta}$ correspond to
the generators of projective transformations. The $15$ generators of
conformal transformations are defined in terms of the set $\left\{
J_{AB}\right\} $ where $A=0$, $1$, $2$,..$5$. The elements $J_{AB}$ can be
decomposed into translations $\mathbf{P}_{\mu}^{\text{Conf}%
}:=J_{5\mu}+J_{6\mu}$, special conformal generators $\mathbf{\Delta}_{\mu
}:=J_{5\mu}-J_{6\mu}$, dilatons $\mathbf{D}:=J_{56}$ and the Lorentz
generators $\mathbf{L}_{\alpha\beta}:=J_{\alpha\beta}$. The Lie algebra of $%
CA(3$, $1)$ is characterized by the commutation relations

\begin{align}
& \left[ \mathbf{\Lambda}_{\alpha\beta}\text{, }\mathbf{D}\right] =\left[ 
\mathbf{\Delta}_{\alpha}\text{, }\mathbf{\Delta}_{\beta}\right] =0\,,  \notag
\\
& \left[ \mathbf{P}_{\alpha}\text{, }\mathbf{P}_{\beta}\right] =\left[ 
\mathbf{D}\text{, }\mathbf{D}\right] = 0\,,  \notag \\
& \left[ \mathbf{L}_{\alpha\beta}\text{, }\mathbf{P}_{\mu}\right]
=io_{\mu\lbrack\alpha}\mathbf{P}_{\beta]}\text{, }\left[ \mathbf{L}%
_{\alpha\beta}\text{, }\mathbf{\Delta}_{\gamma}\right] =io_{[\alpha|\gamma }%
\mathbf{\Delta}_{|\beta]}\text{,}  \notag \\
& \left[ \mathbf{\Lambda}_{\text{ }\beta}^{\alpha}\text{, }\mathbf{P}_{\mu }%
\right] =i\delta_{\mu}^{\alpha}\mathbf{P}_{\beta}\text{, }\left[ \mathbf{%
\Lambda}_{\text{ }\beta}^{\alpha}\text{, }\mathbf{\Delta}_{\mu }\right]
=i\delta_{\mu}^{\alpha}\mathbf{\Delta}_{\beta}\text{,}  \notag \\
& \left[ \mathbf{S}_{\alpha\beta}\text{, }\mathbf{P}_{\mu}\right]
=io_{\mu(\alpha}\mathbf{P}_{\beta)}\text{, }\left[ \mathbf{P}_{\alpha}\text{%
, }\mathbf{D}\right] =-i\mathbf{P}_{\alpha}\text{,}  \notag \\
& \left[ \mathbf{L}_{\alpha\beta}\text{, }\mathbf{L}_{\mu\nu}\right]
=-i\left( o_{\alpha\lbrack\mu}\mathbf{L}_{\nu]\beta}-o_{\beta\lbrack\mu }%
\mathbf{L}_{\nu]\alpha}\right) \,,  \notag \\
& \left[ \mathbf{S}_{\alpha\beta}\text{, }\mathbf{S}_{\mu\nu}\right]
=i\left( o_{\alpha(\mu}\mathbf{L}_{\nu)\beta}-o_{\beta(\mu}\mathbf{L}%
_{\nu)\alpha}\right) \,, \notag\\
& \left[ \mathbf{L}_{\alpha\beta}\text{, }\mathbf{S}_{\mu\nu}\right]
=i\left( o_{\alpha(\mu}\mathbf{S}_{\nu)\beta}-o_{\beta(\mu}\mathbf{S}%
_{\nu)\alpha}\right) \,,  \notag \\
& \left[ \mathbf{\Delta}_{\alpha}\text{, }\mathbf{D}\right] =i\mathbf{\Delta}%
_{\alpha}\text{, }\left[ \mathbf{S}_{\mu\nu}\text{, }\mathbf{\Delta}
_{\alpha}\right] =io_{\alpha(\mu}\mathbf{\Delta}_{\nu )}\,,  \notag \\
& \left[ \mathbf{\Lambda}_{\text{ }\beta}^{\alpha}\text{, }\mathbf{\Lambda }%
_{\text{ }\nu}^{\mu}\right] =i\left( \delta_{\nu}^{\alpha}\mathbf{\Lambda }_{%
\text{ }\beta}^{\mu}-\delta_{\beta}^{\mu}\mathbf{\Lambda}_{\text{ }\nu
}^{\alpha}\right) \,,  \notag \\
& \left[ \mathbf{P}_{\alpha}\text{, }\mathbf{\Delta}_{\beta}\right]
=2i\left( o_{\alpha\beta}\mathbf{D}-\mathbf{L}_{\alpha\beta}\right) \,, 
\end{align} 
where $o_{\alpha\beta}=diag\left( -1\text{, }1\text{, }1\text{, }1\right) $
is the Lorentz group metric. The above algebra is the core of  NLR  and, in some sense, of the invariance induced gravity that we are considering here. The
relation with the $CE$ $sp(8)$ algebra is based on the
observation that the 9 traceless generators of the symmetric linear
transfomations (shear) can be factored in the 8 parameter generators $%
F_{\rho\dot{\tau}}$, $\overline{F}_{\rho\dot{\tau}}\equiv(F_{\tau\dot{\rho}%
})^{\ast}\,$\ plus the axion $A$ \ (see below). As it can be easily seen, the
standard $sp(8)$ algebra has the commutation relations
that are related with the  standard algebra $so(2,4)\simeq su(2,2)$
that is spanned by the generators ($L_{\alpha\dot{\beta}},{\overline{L}}_{%
\dot{\alpha }\dot{\beta}},P_{\alpha\dot{\beta}},K_{\alpha\dot{\beta}},D$), that is 

\begin{align}
& \left[ P_{\alpha\dot{\beta}},P_{\gamma\dot{\delta}}\right] =\left[
K_{\alpha\dot{\beta}},K_{\gamma\dot{\delta}}\right] = 0\,,  \notag \\
\notag \\
& [P_{\alpha\dot{\beta}},K_{\rho\dot{\lambda}}]=\frac{1}{2}\left(
\epsilon_{\alpha\rho}\bar{L}_{\dot{\beta}\dot{\lambda}}-\epsilon_{\dot{\beta 
}\dot{\lambda}}L_{\alpha\rho}\right) -i\epsilon_{\alpha\rho}\epsilon _{\dot{%
\beta}\dot{\lambda}}D\,,  \notag \\
\notag \\
& [L_{\alpha\beta},L_{\rho\lambda}]=\epsilon_{\alpha\rho}L_{\beta\lambda
}+\epsilon_{\beta\rho}L_{\alpha\lambda}+\epsilon_{\alpha\lambda}L_{\beta\rho
}+\epsilon_{\beta\lambda}L_{\rho\alpha}\,,  \notag \\
\notag \\
& [L_{\alpha\beta},P_{\rho\dot{\rho}}]=\epsilon_{\alpha\rho}P_{\beta\dot {%
\rho}}+\epsilon_{\beta\rho}P_{\alpha\dot{\rho}}\,,\ [L_{\alpha\beta},K_{\rho%
\dot{\rho}}]=\epsilon_{\alpha\rho}K_{\beta\dot{\rho}}+\epsilon
_{\beta\rho}K_{\alpha\dot{\rho}}\,,  \notag \\
\notag \\
& [D,P_{\alpha\dot{\alpha}}]=iP_{\alpha\dot{\alpha}}\,,\quad\lbrack
D,K_{\alpha\dot{\alpha}}]=-iK_{\alpha\dot{\alpha}}\,.
\end{align}
The rest of non-vanishing commutators can be obtained by complex conjugation.
The algebra $sl(4,R)$ is spanned by the generators $(L_{\alpha\beta},%
\overline{L}_{\dot{\alpha}\dot{\beta}},A,F_{\alpha\dot{\beta}},\overline {F}%
_{\alpha\dot{\beta}})$. The extra generators $A$, $F_{\rho\dot{\tau}}$, $%
\overline{F}_{\rho\dot{\tau}}\equiv(F_{\tau\dot{\rho}})^{\ast}\,$ satisfy
the relations

\begin{align}
& \left[ A,F_{\alpha\overset{.}{\beta}}\right] =2F_{\alpha\overset{.}{\beta}%
}\,,  \notag \\
& \left[ A,\overline{F}_{\alpha\overset{.}{\beta}}\right] = -2\overline {F}%
_{\alpha\overset{.}{\beta}}\,,  \notag \\
& \left[ F_{\alpha\overset{.}{\beta}},F_{\gamma\overset{.}{\delta}}\right] =%
\left[ \overline{F}_{\alpha\overset{.}{\beta}},\overline{F}_{\gamma \overset{%
.}{\delta}}\right] =0,  \notag \\
& \lbrack F_{\alpha\dot{\beta}},\overline{F}_{\rho\dot{\lambda}}]=2\left(
\epsilon_{\alpha\rho}\bar{L}_{\dot{\beta}\dot{\lambda}}-\epsilon_{\dot{\beta 
}\dot{\lambda}}L_{\alpha\rho}\right) +2\epsilon_{\alpha\rho}\epsilon _{\dot{%
\beta}\dot{\lambda}}A\,.
\end{align}
The generalized $4D$ conformal algebra $sp(8)$ is a closure of the algebras $%
so(2,4)$ and $sl(4,\mathbb{R})\,$. It is obtained by adding to the generators of $%
sl(4,\mathbb{R})$ and the vectorial Abelian translation generators ($P_{\alpha \dot{%
\beta}}, K_{\alpha\dot{\beta}}$) the following additional 12 Abelian
generators

\begin{itemize}
\item ($Z_{\alpha{\beta}}, \overline{Z}_{\dot{\alpha}\dot{\beta}}$)
describing 6 standard tensorial translations

\item (${\widetilde{Z}}_{\alpha\beta}, \overline{{\widetilde{Z}}}_{\dot {%
\alpha}\dot{\beta}}$) describing 6 conformal tensorial translations.
\end{itemize}
Some of the commutation relations that they satisfy are, for example: 
\begin{align}
& \left[ Z_{\alpha\beta},\widetilde{Z}_{\rho\lambda}\right] =\frac{1}{2}%
\left(
\epsilon_{\alpha\rho}L_{\beta\lambda}+\epsilon_{\beta\rho}L_{\alpha\lambda}+%
\epsilon_{\alpha\lambda}L_{\beta\rho}+\epsilon_{\beta
\lambda}L_{\alpha\rho}\right) +\left( \epsilon_{\alpha\rho}\epsilon
_{\beta\lambda}+\epsilon_{\beta\rho}\epsilon_{\alpha\lambda}\right) \left(
iD-\frac{A}{2}\right) \,,  \notag \\
&  \notag \\
& \left[ P_{\alpha\dot{\beta}},\widetilde{Z}_{\rho\lambda}\right] =\frac {1}{%
2}\left( \epsilon_{\alpha\rho}\overline{F}_{\lambda\dot{\beta}%
}+\epsilon_{\alpha\lambda}\overline{F}_{\rho\dot{\beta}}\right) \,,  \notag
\\
&  \notag \\
& \left[ K_{\alpha\dot{\beta}},\widetilde{Z}_{\rho\lambda}\right] =\frac {1}{%
2}\left( \epsilon_{\alpha\rho}F_{\lambda\dot{\beta}}+\epsilon
_{\alpha\lambda}F_{\rho\dot{\beta}}\right) \,,  \notag \\
\end{align}
being the rest described in detail in \cite{ivanov2}. Then, we have the
concrete possibility to make a nonlinear realization of  $Sp(8),$ where
due to the partial isomorphism described before, the conformal affine
structure of the gravitational theories can be included \cite{reviewFOOP}.
As we have  pointed out in the Introduction, the nonlinear realization
of the standard $sp(8)$ algebra described above does not lead to  geometrical
invariants as measures, then we need to perform the nonlinear realization
with the non-standard $CE$ $sp(8)$ algebra. The non-standard algebra
described by the new"hat" generators, defined in  next Section, sets a
new basis where the commutation relations that differs from the standard $%
CE\ sp(8)$ description are 
\begin{align}
& \left[ \widehat{K}_{\alpha\overset{.}{\alpha}},\widehat{K}_{\beta \overset{%
.}{\beta}}\right] =\frac{1}{m^{2}}\left( \epsilon_{\overset{.}{\alpha}%
\overset{.}{\beta}}L_{\alpha\beta}-\epsilon_{\alpha\beta}\overline{L}_{%
\overset{.}{\alpha}\overset{.}{\beta}}\right) \,,  \notag \\
&  \notag \\
& \left[ X_{\alpha\overset{.}{\beta}},X_{\gamma\overset{.}{\delta}}\right]
=-4\left( \epsilon_{\alpha\gamma}\overline{L}_{\overset{.}{\beta}\overset{.}{%
\delta}}-\epsilon_{\overset{.}{\beta}\overset{.}{\delta}}L_{\alpha\gamma
}\right) \,,  \notag \\
&  \notag \\
& [D,\widehat{K}]=-i\left( \widehat{K}+\frac{2P}{m^{2}}\right) \,,  \notag \\
&  \notag \\
& [D,\widehat{\widetilde{Z}}]=-i\left( \widehat{\widetilde{Z}}+\frac {2Z}{%
m^{2}}\right) \,,  \notag \\
&  \notag \\
& \left[ D,A\right] =0,  \notag \\
&  \notag \\
& [A,G]=\left( 2iX\right) \,,  \notag \\
&  \notag \\
& \left[ X,A\right] =2iG\,,  \notag \\
&  \notag \\
& \left[ X_{\alpha\overset{.}{\beta}},G_{\gamma\overset{.}{\delta}}\right]
=-4i\epsilon_{\alpha\gamma}\epsilon_{\overset{.}{\beta}\overset{.}{\delta}%
}A\,,  \notag \\
&  \notag \\
& \left[ A,Z\right] =2Z,  \notag \\
&  \notag \\
& \left[ A,\widetilde{Z}\right] =-2\widetilde{Z},  \notag \\
&  \notag \\
& \left[ A,\widehat{\widetilde{Z}}\right] =-2\left( \widehat{\widetilde {Z}}+%
\frac{2Z}{m^{2}}\right) \,,  \notag \\
& \left[ \widehat{\widetilde{Z}}_{\alpha\beta},\widehat{\widetilde{Z}}%
_{\rho\lambda}\right] =\frac{1}{m^{2}}\left(
\epsilon_{\alpha\rho}L_{\beta\lambda}+\epsilon_{\beta\rho}L_{\alpha\lambda}+%
\epsilon_{\alpha
\lambda}L_{\beta\rho}+\epsilon_{\beta\lambda}L_{\alpha\rho}\right) \,, 
\notag \\
&  \notag \\
& \left[ \widehat{K}_{\alpha\overset{.}{\alpha}},L_{\rho\sigma}\right]
=-\left( \epsilon_{\rho\alpha}\widehat{K}_{\sigma\overset{.}{\alpha}%
}+\epsilon_{\sigma\alpha}\widehat{K}_{\rho\overset{.}{\alpha}}\right) \,, 
\notag \\
&  \notag \\
& \left[ \widehat{K}_{\alpha\overset{.}{\alpha}},\widehat{\widetilde{Z}}%
_{\rho\sigma}\right] =-\frac{1}{2m^{2}}\left( \epsilon_{\rho\alpha }X_{\sigma%
\overset{.}{\alpha}}+\epsilon_{\sigma\alpha}X_{\rho\overset{.}{\alpha}%
}\right) \,,  \notag \\
&  \notag \\
& \left[ X_{\alpha\overset{.}{\alpha}},\widehat{K}_{\beta\overset{.}{\beta}}%
\right] =4\left( \epsilon_{\beta\alpha}\widehat{\widetilde{\overline{Z}}}_{%
\overset{.}{\beta}\overset{.}{\alpha}}+\epsilon_{\overset{.}{\beta}\overset{.%
}{\alpha}}\widehat{\widetilde{Z}}_{\beta\alpha}\right) \,.
\end{align}
In the next Section,  we will see the deep consequences that these new
generators have  to define the geometrical basic invariants of
the theory.


\section{The {\it Non-Linear Realization} approach: the coset field transformations}

\label{cinque} 

As it is well  known, the Maurer-Cartan form for a Lie group G is a distinguished
differential one-form on G that carries the basic infinitesimal information
about the structure of G. As a one-form, the Maurer--Cartan form is peculiar
in that it takes its values in the Lie algebra associated to the Lie group
G. The Lie algebra is identified by the tangent space of G at the
identity. In the context of the NLR approach,  the Cartan
forms play a significant role in order  to define the geometrical
invariants by which the Lagrangian of the theory can  be constructed. As
we will see in detail soon, starting from the algebra associated to the Lie
group G, caracterizing the symmetries of the manifold, we construct (in
general by exponentiation) the respective coset and, by the pullback,  we extract
 univocally the Cartan forms associated to each generator. The Cartan forms give also a geometrical
description of the field dynamics by the implementation of
constraints ({\it e.g.} the "inverse Higgs") to the equations of motion. To
start now with the NLR\ of $Sp(8)$,  we proceed to define our
coset. In the new basis, we introduce the following generators
\begin{equation*}
\begin{array}{ccc}
\widehat{K}_{\alpha\overset{.}{\alpha}}=K_{\alpha\overset{.}{\alpha}}-\frac {%
1}{m^{2}}P_{\alpha\overset{.}{\alpha}}, & \ \ \ \ \widehat{\widetilde{Z}}%
_{\alpha\beta}=\widetilde{Z}_{\alpha\beta}-\frac{1}{m^{2}}Z_{\alpha\beta}, & 
\ \ \ \ \ \widehat{\widetilde{\overline{Z}}}_{\alpha\beta}=\widetilde {Z}%
_{\alpha\beta}-\frac{1}{m^{2}}Z_{\alpha\beta}%
\end{array}%
\end{equation*}%
\begin{equation*}
\begin{array}{cc}
X_{\alpha\overset{.}{\beta}}=F_{\alpha\overset{.}{\beta}}+\overline{F}%
_{\alpha\overset{.}{\beta}}, & \ \ G_{\alpha\overset{.}{\beta}}=i\left(
F_{\alpha\overset{.}{\beta}}-\overline{F}_{\alpha\overset{.}{\beta}}\right) ,%
\end{array}%
\end{equation*}
where the coset is rewritten as%
\begin{equation*}
g=e^{ix\cdot P}e^{iz\cdot Z}e^{i\phi D}e^{i\alpha A}e^{i\mu\cdot X}e^{ik\cdot%
\widehat{K}}e^{it\cdot\widehat{\widetilde{Z}}}\,.
\end{equation*}
This particular factorization of the coset is convenient in order to have
non-trivial field interactions in the dynamical equations,  once the
Lagrangian of the theory is constructed by the Cartan forms. Then, the
left covariant Cartan forms are extracted from the well know pullback
equation 
\begin{equation*}
-ig^{-1}dg=\omega_{P}\cdot P+\omega_{Z}\cdot Z+\omega_{D}D+\omega_{X}\cdot
X+\omega_{\widehat{K}}\cdot\widehat{K}+\omega_{\widehat{\widetilde{Z}}}\cdot%
\widehat{\widetilde{Z}}+\omega_{A}A+\omega_{G}\cdot G+\omega_{L}\cdot L\,.
\end{equation*}
Notice that:

\begin{itemize}
\item Generators $G_{\alpha\overset{.}{\beta}}$ besides $L_{\alpha\beta
}\left( \overline{L}_{\overset{.}{\alpha}\overset{.}{\beta}}\right) $ are in
the stability group (and not only the Lorentz generators, as is in the
standard NLR approach applied to gravity).

\item It is useful to remark the fact that other combinations of generators of the
algebra can form part of the above stability subgroup, but the parameters
(fields) related to generators $X_{\alpha\overset{.}{\beta}}$  have a
trivial role in the dynamics and interactions. In this sense, this non-uniqueness 
of the factorization (parametrization) of the group elements  in the coset is a problem that can only be solved by  physical
interpretation. Theoretically, this fact goes beyond the non-uniqueness (almost
at  the classical level) of the Lagrangian functions, and it is deeply associated
to the relation between the constraint elimination of the Goldstone fields
from the Cartan forms and the algebraic (non dynamical ) equation of motion
for the respective parameters from the  corresponding Lagrangian function.
\end{itemize}


\section{The Cartan Forms}

\label{sei} 

Now we proceed to extract explicitly, from the pullback equation of G defined
in the previous Section,  the relevant Cartan forms in order to obtain the
building blocks by which we can construct the possible geometrical
invariants that can play the  role of the Lagrangian. They are the following
\begin{eqnarray}
\omega_{D}&=&i\frac{1+\frac{\lambda^{2}}{2}}{1-\frac{\lambda^{2}}{2}}fd\phi-%
\frac{d\alpha}{2}\frac{1+\eta^{2}}{1-\eta^{2}}f--\frac{e^{\phi}}{1-\eta^{2}}%
\left( \frac{\Sigma^{\alpha\beta}\xi_{\alpha\beta}}{1-\frac {\xi^{2}}{2}}+%
\frac{\overline{\Sigma}^{\overset{\cdot}{\alpha}\overset{.}{\beta}}\overline{%
\xi}_{\overset{\cdot}{\alpha}\overset{.}{\beta}}}{1-\frac{\overline{\xi}^{2}%
}{2}}\right)  \notag \\
&& -\frac{e^{\phi}}{1-\eta^{2}}\frac{imf\lambda_{\alpha\overset{.}{\alpha}%
}W^{\alpha\overset{\cdot}{\alpha}}}{1-\frac{\lambda^{2}}{2}},
\end{eqnarray}

\begin{eqnarray}
-i\omega_{A} & =&-i\frac{1+\frac{\lambda^{2}}{2}}{1-\frac{\lambda^{2}}{2}}%
fd\phi+\frac{d\alpha}{2}\frac{1+\eta^{2}}{1-\eta^{2}}f- -\frac{e^{\phi}}{%
1-\eta^{2}}\left( \frac{\Sigma^{\alpha\beta}\xi_{\alpha\beta}}{1-\frac {%
\xi^{2}}{2}}+\frac{\overline{\Sigma}^{\overset{\cdot}{\alpha}\overset{.}{%
\beta}}\overline{\xi}_{\overset{\cdot}{\alpha}\overset{.}{\beta}}}{1-\frac{%
\overline{\xi}^{2}}{2}}\right)   \notag \\
&& +\frac{e^{\phi}}{1-\eta^{2}}\frac{imf\lambda_{\alpha\overset{.}{\alpha}%
}W^{\alpha\overset{\cdot}{\alpha}}}{1-\frac{\lambda^{2}}{2}},
\end{eqnarray}

\begin{eqnarray}
\omega_{P}^{\gamma\overset{.}{\gamma}}=\frac{1}{2}\left[ \frac{-2fd\phi }{1-%
\frac{\lambda^{2}}{2}}\lambda^{\gamma\overset{\cdot}{\gamma}}+\frac{%
me^{\phi}f}{1-\eta^{2}}\left( \delta_{\alpha}^{\gamma}\delta _{\overset{\cdot%
}{\alpha}}^{\overset{\cdot}{\gamma}}+\frac{\lambda _{\alpha\overset{\cdot}{%
\alpha}}\lambda^{\beta\overset{\cdot}{\beta}}}{1-\frac{\lambda^{2}}{2}}%
\right) W^{\alpha\overset{\cdot}{\alpha}}\right] ,
\end{eqnarray}

\begin{eqnarray}
\omega_{\widehat{K}}^{\rho\overset{.}{\rho}} & =&\frac{1}{2}\left[ f\frac{%
d\lambda^{\gamma\overset{\cdot}{\gamma}}}{1-\frac{\lambda^{2}}{2}}-\frac{%
2fmd\phi}{1-\frac{\lambda^{2}}{2}}\lambda^{\gamma\overset{\cdot }{\gamma}}-%
\frac{m^{2}e^{\phi}g}{1-\eta^{2}}\left( \delta_{\alpha}^{\gamma }\delta_{%
\overset{\cdot}{\alpha}}^{\overset{\cdot}{\gamma}}+\frac {\lambda_{\alpha%
\overset{\cdot}{\alpha}}\lambda^{\beta\overset{\cdot}{\beta}}}{1-\frac{%
\lambda^{2}}{2}}\right) W^{\alpha\overset{\cdot}{\alpha}}+\frac{2mg\ d\phi}{%
1-\frac{\lambda^{2}}{2}}\lambda^{\gamma\overset{\cdot }{\gamma}}\right. 
\notag \\
&& \left. -\frac{m^{2}e^{\phi}f}{1-\eta^{2}}\left( \frac{-2\lambda _{\alpha%
\overset{\cdot}{\alpha}}\lambda^{\gamma\overset{.}{\gamma}%
}+\lambda^{2}\delta_{\alpha}^{\gamma}\delta_{\overset{\cdot}{\alpha}}^{%
\overset{\cdot}{\gamma}}}{1-\frac{\lambda^{2}}{2}}\right) W^{\alpha \overset{%
\cdot}{\alpha}}\right] -\frac{d\eta^{\rho\overset{.}{\rho}}iP_{\rho\overset{.%
}{\rho}}^{\gamma\overset{.}{\gamma}}}{1-\eta^{2}}\frac{1+\frac{\lambda^{2}}{2%
}}{1-\frac{\lambda^{2}}{2}}  \notag \\
&& = \frac{1}{2}f\left[ \frac{d\lambda^{\gamma\overset{\cdot}{\gamma}}}{1-%
\frac{\lambda^{2}}{2}}-\frac{2md\phi}{1-\frac{\lambda^{2}}{2}}\lambda^{\gamma%
\overset{\cdot}{\gamma}}-\frac{m^{2}e^{\phi}}{1-\eta^{2}}\left( \frac{%
-2\lambda_{\alpha\overset{\cdot}{\alpha}}\lambda^{\gamma \overset{.}{\gamma}%
}+\lambda^{2}\delta_{\alpha}^{\gamma}\delta_{\overset{\cdot}{\alpha}}^{%
\overset{\cdot}{\gamma}}}{1-\frac{\lambda^{2}}{2}}\right) W^{\alpha\overset{%
\cdot}{\alpha}}\right] -\frac{d\eta^{\rho\overset{.}{\rho}}iP_{\rho\overset{.%
}{\rho}}^{\gamma\overset{.}{\gamma}}}{1-\eta^{2}}\,,  \notag \\
\end{eqnarray}

\begin{eqnarray}
\omega_{Z}^{\alpha\beta} & =&\frac{1}{2}\left[ \frac{d\phi}{m}\frac {1-\frac{%
\lambda^{2}}{2}}{1-\frac{\lambda^{2}}{2}}-\frac{e^{\phi}}{1-\eta^{2}}\frac{%
\lambda_{\alpha\overset{.}{\alpha}}W^{\alpha\overset{\cdot}{\alpha}}}{1-%
\frac{\lambda^{2}}{2}}\right] \left[ \frac{\xi^{\alpha\beta}}{1-\frac{\xi^{2}%
}{2}}-\frac{\eta^{\ \alpha\overset{.}{\alpha}}\lambda _{\ \ \overset{.}{%
\alpha}}^{\beta}}{\left( 1+\eta^{2}\right) \left( 1-\frac{\lambda^{2}}{2}%
\right) }\right]   \notag \\
&&+ 2d\alpha\frac{1+\eta^{2}}{1-\eta^{2}}\frac{i}{m}\frac{\xi^{\alpha\beta}}{%
1-\frac{\xi^{2}}{2}} +i\frac{e^{\phi}}{1-\eta^{2}}\left(
\delta_{\gamma}^{\alpha}\delta _{\delta}^{\beta}+\frac{\xi^{\alpha\beta}%
\xi_{\gamma\delta}}{1-\frac{\xi^{2}}{2}}\right) \Sigma^{\gamma\delta}\,,
\end{eqnarray}

\begin{eqnarray}
\omega_{\widehat{\widetilde{Z}}}^{\alpha\beta}=2m^{2}\omega_{Z}^{\alpha\beta
}-i\frac{d\eta^{\alpha\overset{\cdot}{\rho}}}{1-\eta^{2}}\frac{\lambda _{%
\overset{.}{\rho}}^{\beta}}{1-\frac{\lambda^{2}}{2}}+\frac{1+\frac {%
\lambda^{2}}{2}}{1-\frac{\lambda^{2}}{2}}\frac{d\xi^{\alpha\beta}}{1-\frac{%
\xi^{2}}{2}}\,,
\end{eqnarray}

\begin{eqnarray}
\omega_{G}^{\rho\overset{.}{\rho}} & =&\left[ \frac{-2md\phi}{1-\frac {%
\lambda^{2}}{2}}\lambda^{\gamma\overset{\cdot}{\gamma}}+\frac{m^{2}e^{\phi}}{%
1-\eta^{2}}\left( \delta_{\alpha}^{\gamma}\delta_{\overset{\cdot}{\alpha}}^{%
\overset{\cdot}{\gamma}}+\frac{\lambda_{\alpha\overset{\cdot}{\alpha}%
}\lambda^{\gamma\overset{\cdot}{\gamma}}}{1-\frac{\lambda^{2}}{2}}\right)
W^{\alpha\overset{\cdot}{\alpha}}\right] iP_{\gamma\overset{.}{\gamma}}^{\rho%
\overset{.}{\rho}}  \notag \\
&& +\frac{e^{\phi}}{1-\eta^{2}}\left( \frac {\Sigma^{\alpha\rho}\lambda_{%
\alpha\ \ }^{\overset{\cdot}{\ \rho}}}{\left( 1-\frac{\lambda^{2}}{2}\right) 
}+\frac{\overline{\Sigma}^{\overset{\cdot }{\alpha}\overset{.}{\rho}%
}\lambda_{\ \ \overset{.}{\alpha}}^{\rho}}{\left( 1-\frac{\lambda^{2}}{2}%
\right) }\right) +fd\alpha\frac{\eta^{\gamma\overset{\cdot}{\gamma}}}{%
1-\eta^{2}}\frac{1+\frac{\lambda^{2}}{2}}{1-\frac{\lambda^{2}}{2}}\,,
\end{eqnarray}

\begin{eqnarray}
\omega_{X}^{\rho\overset{.}{\rho}} & =\frac{1}{1-\eta^{2}}\left[ -\frac{%
d\lambda^{\gamma\overset{\cdot}{\gamma}}}{1-\frac{\lambda^{2}}{2}}%
+e^{\phi}\left( \frac{-2\lambda_{\alpha\overset{\cdot}{\alpha}%
}\lambda^{\gamma\overset{.}{\gamma}}+\lambda^{2}\delta_{\alpha}^{\gamma}%
\delta_{\overset{\cdot}{\alpha}}^{\overset{\cdot}{\gamma}}}{1-\frac {%
\lambda^{2}}{2}}\right) W^{\alpha\overset{\cdot}{\alpha}}+\frac{2d\phi }{1-%
\frac{\lambda^{2}}{2}}\lambda^{\gamma\overset{\cdot}{\gamma}}\right]
iP_{\gamma\overset{.}{\gamma}}^{\rho\overset{.}{\rho}}  \notag \\
& +f\frac{d\eta^{\rho\overset{\cdot}{\rho}}}{1-\eta^{2}}\frac{1+\frac {%
\lambda^{2}}{2}}{1-\frac{\lambda^{2}}{2}}-\frac{1}{1-\frac{\lambda^{2}}{2}}%
\left( \frac{d\xi^{\alpha\rho}\lambda_{\alpha\ \ }^{\overset{\cdot}{\ \rho}}%
}{1-\frac{\xi^{2}}{2}}+\frac{d\overline{\xi}^{\overset{\cdot}{\alpha}\overset%
{.}{\rho}}\lambda_{\ \ \overset{.}{\alpha}}^{\rho}}{\left( 1-\frac{\overline{%
\xi}^{2}}{2}\right) }\right) \,,
\end{eqnarray}
where we define the following important quantities:%
\begin{eqnarray}
f\equiv\frac{1+\frac{\xi^{2}}{2}}{1-\frac{\xi^{2}}{2}}+\frac{1+\frac {%
\overline{\xi}^{2}}{2}}{1-\frac{\overline{\xi}^{2}}{2}}\text{\ , \ \ \ \ \ \ 
}g\equiv\frac{\xi^{2}}{1-\frac{\xi^{2}}{2}}+\frac{\overline{\xi }^{2}}{1-%
\frac{\overline{\xi}^{2}}{2}}\,,
\end{eqnarray}
the 1-forms%
\begin{eqnarray}
W^{\alpha\overset{\cdot}{\alpha}}\equiv\left( 1+\eta^{2}\right) dx^{\alpha%
\overset{.}{\alpha}}+2i\left( e^{-2i\alpha}dz^{\sigma\beta}\eta_{\sigma}^{\
\ \overset{.}{\alpha}}+e^{2i\alpha}d\overset{\_}{z}^{\overset{.}{\sigma}%
\overset{.}{\alpha}}\eta_{\ \ \overset{.}{\sigma}}^{\alpha}\right) \,,
\end{eqnarray}
\begin{eqnarray}
\Sigma^{\alpha\beta}\equiv\left[ \eta_{\ \ \overset{.}{\alpha}}^{\beta
}dx^{\alpha\overset{.}{\alpha}}+ie^{-2i\alpha}\left( 1+\eta^{2}\right)
dz^{\alpha\beta}\right]
\end{eqnarray}
\begin{eqnarray}
\overline{\Sigma}^{\overset{.}{\alpha}\overset{.}{\beta}}\equiv\left[
\eta_{\alpha}^{\ \ \overset{.}{\beta}}dx^{\alpha\overset{.}{\alpha}%
}-ie^{+2i\alpha}\left( 1+\eta^{2}\right) d\overset{\_}{z}^{\overset{.}{\alpha%
}\overset{.}{\beta}}\right]
\end{eqnarray}
the associated vectors in the dual space fulfilling the relations 
\begin{eqnarray}
\left\langle W^{\alpha\overset{\cdot}{\alpha}},\left( W^{-1}\right) _{\alpha%
\overset{\cdot}{\alpha}}\right\rangle =\delta_{\text{ }\beta \overset{\cdot}{%
\beta}}^{\alpha\overset{\cdot}{\alpha}};\left\langle
\Sigma^{\alpha\beta},\left( \Sigma^{-1}\right) _{\rho\sigma}\right\rangle
=\delta_{\text{ }\rho\sigma}^{\alpha\beta}\,,
\end{eqnarray}

\begin{eqnarray}
\left\langle \overline{\Sigma}^{\overset{.}{\alpha}\overset{.}{\beta}%
},\left( \overline{\Sigma}^{-1}\right) _{\overset{.}{\rho}\overset{.}{\sigma}%
}\right\rangle =\delta_{\text{ }\overset{.}{\rho}\overset{.}{\sigma}}^{%
\overset{.}{\alpha}\overset{.}{\beta}}\,,
\end{eqnarray}

\begin{eqnarray}
\left\langle W^{\alpha\overset{\cdot}{\alpha}},\left( \Sigma^{-1}\right)
_{\alpha\beta}\right\rangle =\left\langle W^{\alpha\overset{\cdot}{\alpha}%
},\left( \overline{\Sigma}^{-1}\right) _{\overset{.}{\alpha}\overset{.}{\beta%
}}\right\rangle =\left\langle \Sigma^{\alpha\beta},\left( \overline{\Sigma}%
^{-1}\right) _{\overset{.}{\alpha}\overset{.}{\beta}}\right\rangle =0\,,
\end{eqnarray}

\begin{align}
\left( W^{-1}\right) _{\alpha\overset{\cdot}{\alpha}}\equiv\frac{\left(
1+\eta^{2}\right) }{\Delta}\left[ \left( 1+\eta^{2}\right) \partial _{\alpha%
\overset{\cdot}{\alpha}}-2e^{2i\alpha}\left( \eta_{\alpha }^{\ \ \overset{.}{%
\beta}}\partial_{\overset{\cdot}{\alpha}\overset{\cdot }{\beta}}-\eta_{\ \ 
\overset{.}{\alpha}}^{\beta}\partial_{\alpha\beta}\right) \right] \,,
\end{align}%
\begin{align}
\left( \Sigma^{-1}\right) _{\alpha\beta}\equiv\frac{-ie^{2i\alpha}}{\Delta }%
\left[ \left( \left( 1+\eta^{2}\right) ^{2}-2\eta^{2}\right)
\partial_{\alpha\beta}-\left( 1+\eta^{2}\right) \eta_{\beta}^{\ \ \overset{.}%
{\alpha}}\partial_{\alpha\overset{\cdot}{\alpha}}-2\eta_{\beta }^{\ \ 
\overset{.}{\alpha}}\eta_{\alpha}^{\ \ \overset{.}{\beta}}\partial_{\overset{%
\cdot}{\alpha}\overset{\cdot}{\beta}}\right] \,,
\end{align}
\begin{align}
\left( \overline{\Sigma}^{-1}\right) _{\overset{.}{\alpha}\overset{.}{\beta }%
}\equiv\frac{ie^{-2i\alpha}}{\Delta}\left[ \left( \left( 1+\eta ^{2}\right)
^{2}+2\eta^{2}\right) \partial_{\overset{\cdot}{\alpha}\overset{\cdot}{\beta}%
}+2\eta_{\ \ \overset{.}{\beta}}^{\alpha}\eta _{\ \ \overset{.}{\alpha}%
}^{\beta}\partial_{\alpha\beta}-\left( 1+\eta ^{2}\right) \eta_{\ \ \overset{%
.}{\beta}}^{\alpha}\partial_{\alpha \overset{\cdot}{\alpha}}\right] \,,
\end{align}

with%
\begin{equation*}
\Delta=\left( 1+\eta^{2}\right) ^{10}\,.
\end{equation*}
These one forms and the corresponding (dual) vectors are directly related
with the gauge covariant basis in which the metric splits in several blocks
corresponding to the Manifold subgroups. These subgroups represent (as we
will see later) the space-time  and the additional symmetries. We 
have also the following projector%
\begin{equation*}
P_{\rho\overset{.}{\rho}}^{\gamma\overset{.}{\gamma}}=\frac{%
\xi_{\rho}^{\gamma}\delta_{\overset{.}{\rho}}^{\overset{.}{\gamma}}}{1-\frac{%
\xi^{2}}{2}}+\frac{\overline{\xi}_{\overset{.}{\rho}}^{\overset{.}{\gamma}%
}\delta_{\rho}^{\gamma}}{1-\frac{\overline{\xi}^{2}}{2}}\,.
\end{equation*}
The standard transformations on the parameters in  to convert
hyperbolic and trigonometrical functions in polynomial ones are
\begin{equation*}
\lambda^{\rho\overset{\cdot}{\sigma}}=\frac{tanh\sqrt{\frac{k^{2}}{2m^{2}}}}{%
\sqrt{\frac{k^{2}}{2m^{2}}}}\left(\frac{k^{\rho\overset{\cdot}{\sigma}}}{m}\right),\ \ \
\ \ \ \eta^{\rho\overset{.}{\sigma}}=\frac{tanh\sqrt{\mu^{2}}}{\sqrt{\mu^{2}}%
}\mu^{\rho\overset{.}{\sigma}}\,,
\end{equation*}%
\begin{equation*}
\xi^{\rho\sigma}=\frac{tanh\sqrt{\frac{t^{2}}{m^{2}}}}{\sqrt{\frac{t^{2}}{%
m^{2}}}}\left(\frac{t^{\rho\sigma}}{m}\right)\ ,\ \ \ \ \ \ \overline{\xi}^{\overset{\cdot%
}{\rho}\overset{\cdot}{\sigma}}=\frac{tanh\sqrt{\frac{\overset{\_}{t}^{2}}{%
m^{2}}}}{\sqrt{\frac{\overset{\_}{t}^{2}}{m^{2}}}}\left(\frac{\overline {t}^{%
\overset{\cdot}{\rho}\overset{\cdot}{\sigma}}}{m}\right)
\end{equation*}
simplifying considerabily the theoretical analysis.


\section{Inverse Higgs constraints}

\label{sette} 

In order to eliminate $\lambda^{\rho\overset{\cdot}{\sigma}},$ $\xi
^{\rho\sigma}$ in a covariant suitable form,  we have to take into account  a particular
simplification. Such a  simplification is based on the separability condition
coming from the same structure of the $Sp(8)$ group: that is, if
we impose $\alpha=\alpha\left( \xi_{\alpha\beta}\right) $ and $\phi
=\phi\left( \lambda_{\alpha\overset{.}{\alpha}}\right),$ the consistent
inverse Higgs constraint $\omega_{D}=0$ leads to the simple equations: 
\begin{equation*}
d\phi=\frac{e^{\phi}}{1-\eta^{2}}\left(\frac{m\lambda_{\alpha\overset{.}{\alpha}%
}W^{\alpha\overset{\cdot}{\alpha}}}{1+\frac{\lambda^{2}}{2}}\right)\,,
\end{equation*}
and%
\begin{equation*}
-d\alpha=\frac{me^{\phi}}{1+\eta^{2}}\left(\frac{\Sigma^{\alpha\beta}\xi
_{\alpha\beta}}{1+\frac{\xi^{2}}{2}}\right)\,,
\end{equation*}
where the underlying role played by the axion and dilaton with respect to
the group manifold under consideration is quite evident. Then, taking into
account the above conditions due to the inverse Higgs constraint, both $%
\lambda^{\rho \overset{\cdot}{\sigma}},$ $\xi^{\rho\sigma}$ can be easily
separately eliminated. It is not so simple in the general case where $\phi$ and $%
\alpha$ depend both on $\lambda_{\alpha\overset{.}{\alpha}}$ and $%
\xi_{\alpha\beta}$. For $\lambda^{\rho\overset{\cdot}{\sigma}}$, we have 
\begin{eqnarray}
\overset{\equiv\chi_{1}}{\overbrace{\frac{me^{\phi}}{\left( 1-\eta
^{2}\right) }}}\frac{\lambda_{\alpha\overset{.}{\alpha}}}{\left( 1+\frac{%
\lambda^{2}}{2}\right) }=\left( W^{-1}\right) _{\alpha \overset{\cdot}{\alpha%
}}\phi\,, \quad\Rightarrow\quad\lambda_{\alpha \overset{.}{\alpha}}=\frac{%
\left( W^{-1}\right) _{\alpha\overset{\cdot }{\alpha}}\phi}{\chi_{1}\left( 1+%
\sqrt{1-\frac{2\left( W^{-1}\right) _{\alpha\overset{.}{\alpha}}\phi\left(
W^{-1}\right) ^{\alpha\overset{.}{\alpha}}\phi}{\chi_{1}^{2}}}\right) }\,. 
\notag \\
\end{eqnarray}
Similarly, for $\xi_{\alpha\beta}$ $\left( \overline{\xi}_{\overset{.}{\alpha 
}\overset{.}{\beta}}\right) $, one has

\begin{eqnarray}
-\frac{d\alpha}{2}=\overset{\equiv\chi_{2}}{\overbrace{\frac{me^{\phi}}{%
1-\eta^{2}}}}\frac{\Sigma^{\alpha\beta}\xi_{\alpha\beta}}{1+\frac{\xi^{2}}{2}%
}\,, \qquad\Rightarrow\qquad\xi_{\alpha\beta}=\frac{\left( \Sigma
^{-1}\right) _{\alpha\beta}\alpha}{\chi_{2}\left( 1+\sqrt{1-\frac{2\left(
\Sigma^{-1}\right) _{\alpha\beta}\alpha\left( \Sigma^{-1}\right)
^{\alpha\beta}\alpha}{\chi_{2}^{2}}}\right) }\,.  \notag \\
\end{eqnarray}
However, $\overline{\xi}_{\overset{.}{\alpha}\overset{.}{\beta}}$ is
eliminated in an analog manner.
On the surface of this covariant constraint, the remaining coset Cartan forms
are given by the expressions%
\begin{eqnarray}
\left. \omega_{A}\right\vert _{\omega_{D}=0}=i\frac{me^{\phi}}{1-\eta^{2}}%
\left( \frac{\Sigma^{\alpha\beta}\xi_{\alpha\beta}}{1-\frac{\xi^{2}}{2}}+%
\frac{\overline{\Sigma}^{\overset{\cdot}{\alpha}\overset{.}{\beta}}\overline{%
\xi}_{\overset{\cdot}{\alpha}\overset{.}{\beta}}}{1-\frac {\overline{\xi}^{2}%
}{2}}\right) ,
\end{eqnarray}
\begin{eqnarray}
\left. \omega_{P}^{\rho\overset{.}{\rho}}\right\vert
_{\omega_{D}=0}=\chi\left( \delta_{\alpha}^{\rho}\delta_{\overset{.}{\alpha}%
}^{\overset{.}{\rho}}-\frac{\lambda_{\alpha\overset{.}{\alpha}}\lambda^{\rho%
\overset{.}{\rho}}}{1+\frac{\lambda^{2}}{2}}\right) W^{\alpha\overset{\cdot}{%
\alpha}}\equiv E_{\alpha\overset{.}{\alpha}}^{\rho\overset{.}{\rho}%
}W^{\alpha \overset{\cdot}{\alpha}}=e^{\phi}\widehat{E}_{\alpha\overset{.}{%
\alpha}}^{\rho\overset{.}{\rho}}W^{\alpha\overset{\cdot}{\alpha}},
\end{eqnarray}
\begin{eqnarray}
\left. \omega_{\widehat{K}}^{\rho\overset{.}{\rho}}\right\vert _{\omega
_{D}=0} & =&\frac{fd\lambda^{\rho\overset{.}{\rho}}}{1-\frac{\lambda^{2}}{2}}%
-\frac{B\ \chi m}{2\left( 1-\frac{\lambda^{2}}{2}\right) }\left(
\delta_{\alpha}^{\rho}\delta_{\overset{.}{\alpha}}^{\overset{.}{\rho}}-\frac{%
\lambda_{\alpha\overset{.}{\alpha}}\lambda^{\rho\overset{.}{\rho}}}{1+\frac{%
\lambda^{2}}{2}}\right) W^{\alpha\overset{\cdot}{\alpha}}+\frac{d\eta^{\gamma%
\overset{.}{\gamma}}P_{\gamma\overset{.}{\gamma}}^{\rho\overset{.}{\rho}}}{%
1-\eta^{2}}  \notag \\
&& =\frac{1}{1-\frac{\lambda^{2}}{2}}\left( fd\lambda^{\rho\overset{.}{\rho}%
}-\frac{\lambda^{2}+\xi^{2}+\overline{\xi}^{2}}{1-\left( \lambda^{2}+\xi
^{2}+\overline{\xi}^{2}\right) }\omega_{P}^{\rho\overset{.}{\rho}}\right) +%
\frac{d\eta^{\gamma\overset{.}{\gamma}}P_{\gamma\overset{.}{\gamma}}^{\rho%
\overset{.}{\rho}}}{1-\eta^{2}},
\end{eqnarray}

\begin{eqnarray}
\left. \omega_{Z}^{\alpha\beta}\right\vert _{\omega_{D}=0}=i\frac{e^{\phi}}{%
1-\eta^{2}}\left( \delta_{\gamma}^{\alpha}\delta_{\delta}^{\beta}-\frac {%
\xi^{\alpha\beta}\xi_{\gamma\delta}}{1+\frac{\xi^{2}}{2}}\right)
\Sigma^{\gamma\delta},
\end{eqnarray}
\begin{eqnarray}
\left. \omega_{\widehat{\widetilde{Z}}}^{\alpha\beta}\right\vert _{\omega
_{D}=0}=\frac{1+\frac{\lambda^{2}}{2}}{1-\frac{\lambda^{2}}{2}}\frac {%
d\xi^{\alpha\beta}}{1-\frac{\xi^{2}}{2}}-i\frac{d\eta^{\alpha\overset{\cdot }%
{\rho}}}{1-\eta^{2}}\frac{\lambda_{\overset{.}{\rho}}^{\beta}}{1-\frac {%
\lambda^{2}}{2}}+2m^{2}\omega_{Z}^{\alpha\beta},
\end{eqnarray}
\begin{eqnarray}
\left. \omega_{G}^{\rho\overset{.}{\rho}}\right\vert
_{\omega_{D}=0}=im\omega_{P}^{\gamma\overset{\cdot}{\gamma}}P_{\gamma\overset%
{.}{\gamma}}^{\rho\overset{.}{\rho}}-if\frac{\eta^{\gamma\overset{\cdot}{%
\gamma}}}{1+\eta^{2}}\frac{1+\frac{\lambda^{2}}{2}}{1-\frac{\lambda^{2}}{2}}%
\omega _{A},
\end{eqnarray}
\begin{eqnarray}
\left. \omega_{X}^{\rho\overset{.}{\rho}}\right\vert _{\omega_{D}=0}=-\frac{%
iP_{\gamma\overset{.}{\gamma}}^{\rho\overset{.}{\rho}}}{\left( 1-\frac{%
\lambda^{2}}{2}\right) }\left[ \frac{d\lambda^{\gamma\overset{\cdot }{\gamma}%
}}{\left( 1-\eta^{2}\right) }-\lambda^{2}m\omega_{P}^{\gamma \overset{\cdot}{%
\gamma}}\right] +f\frac{d\eta^{\rho\overset{\cdot}{\rho}}}{1-\eta^{2}},
\end{eqnarray}
where we have defined 
\begin{equation}
B=\left( -2+f\frac{1+\lambda^{2}/2}{1-\lambda^{2}/2}\right)\,.
\end{equation}
Notice that the geometrical role of $\eta^{\rho\overset{\cdot}{\rho}},$
namely, as it enter in the conformal Cartan forms$\left. \omega_{\widehat{K}%
}^{\rho\overset{.}{\rho}}\right\vert _{\omega_{D}=0}$ and $\left. \omega_{%
\widehat{\widetilde{Z}}}^{\alpha\beta}\right\vert _{\omega_{D}=0}$and the
action as projector-connection in $W^{\alpha\overset{\cdot}{\alpha}}$ and $%
\Sigma^{\alpha\beta}$ , are intrinsically related with its role into the
gauge potentials as we will show in detail.

. 

\section{The G-Manifold, spacetime and affine structure}

\label{otto} 

We have  seen that, after the breaking of the conformal symmetry ($%
\omega_{D}=0)$, the original group manifold plays the role of a 
10-dimensional space-time, spanned by W, $\Sigma$ and $\overline{\Sigma}.$ 
Once the Higgs (inverse-constraint) mechanism is established, the 10-dimensional spacetime (after a suitable choice of the Lagrangian) is
transformed in the arena where $\eta,\alpha,$and $\phi$ are the dynamical fields of
the theory. Notice that when $\xi^{\rho\sigma}$ and $\eta_{\sigma}^{\ \ 
\overset{.}{\alpha}}$ are null, the well-known pure conformal case is
recovered.

Previously, we have  shown that there exist a basis where the structure
of the Cartan forms (derived by the pullback process in the nonlinear
realization of $CE$ $Sp(8)$ is clearly seen. The basis is given by the
1-forms%
\begin{align}
& W^{\alpha\overset{\cdot}{\alpha}}\equiv\left( 1+\eta^{2}\right) dx^{\alpha%
\overset{.}{\alpha}}+2i\left( e^{-2i\alpha}dz^{\sigma\beta}\eta_{\sigma}^{\
\ \overset{.}{\alpha}}+e^{2i\alpha}d\overset{\_}{z}^{\overset{.}{\sigma}%
\overset{.}{\alpha}}\eta_{\ \ \overset{.}{\sigma}}^{\alpha}\right) \,, 
\notag \\
&  \notag \\
& \Sigma^{\alpha\beta}\equiv\left[ \eta_{\ \ \overset{.}{\alpha}}^{\beta
}dx^{\alpha\overset{.}{\alpha}}+ie^{-2i\alpha}\left( 1+\eta^{2}\right)
dz^{\alpha\beta}\right] \,,  \notag \\
&  \notag \\
& \overline{\Sigma}^{\overset{.}{\alpha}\overset{.}{\beta}}\equiv\left[
\eta_{\alpha}^{\ \ \overset{.}{\beta}}dx^{\alpha\overset{.}{\alpha}%
}-ie^{+2i\alpha}\left( 1+\eta^{2}\right) d\overset{\_}{z}^{\overset{.}{\alpha%
}\overset{.}{\beta}}\right] \,.
\end{align}
The associated vectors in the dual space%
\begin{align}
& \left( W^{-1}\right) _{\alpha\overset{\cdot}{\alpha}}\equiv\frac{\left(
1+\eta^{2}\right) }{\Delta}\left[ \left( 1+\eta^{2}\right) \partial _{\alpha%
\overset{\cdot}{\alpha}}-2e^{2i\alpha}\left( \eta_{\alpha }^{\ \ \overset{.}{%
\beta}}\partial_{\overset{\cdot}{\alpha}\overset{\cdot }{\beta}}-\eta_{\ \ 
\overset{.}{\alpha}}^{\beta}\partial_{\alpha\beta}\right) \right] \,,  \notag
\\
&  \notag \\
& \left( \Sigma^{-1}\right) _{\alpha\beta}\equiv\frac{-ie^{2i\alpha}}{\Delta}%
\left[ \left( \left( 1+\eta^{2}\right) ^{2}-2\eta^{2}\right)
\partial_{\alpha\beta}-\left( 1+\eta^{2}\right) \eta_{\beta}^{\ \ \overset{.}%
{\alpha}}\partial_{\alpha\overset{\cdot}{\alpha}}-2\eta_{\beta }^{\ \ 
\overset{.}{\alpha}}\eta_{\alpha}^{\ \ \overset{.}{\beta}}\partial_{\overset{%
\cdot}{\alpha}\overset{\cdot}{\beta}}\right] \,,  \notag \\
&  \notag \\
& \left( \overline{\Sigma}^{-1}\right) _{\overset{.}{\alpha}\overset{.}{\beta%
}}\equiv\frac{ie^{-2i\alpha}}{\Delta}\left[ \left( \left( 1+\eta^{2}\right)
^{2}+2\eta^{2}\right) \partial_{\overset{\cdot}{\alpha }\overset{\cdot}{\beta%
}}+2\eta_{\ \ \overset{.}{\beta}}^{\alpha}\eta_{\ \ \overset{.}{\alpha}%
}^{\beta}\partial_{\alpha\beta}-\left( 1+\eta^{2}\right) \eta_{\ \ \overset{.%
}{\beta}}^{\alpha}\partial _{\alpha\overset{\cdot}{\alpha}}\right] \,,
\end{align}
with%
\begin{equation*}
\Delta=\left( 1+\eta^{2}\right) ^{10}\,.
\end{equation*}
These bases of vectors and forms are extremely important: they are the gauge
covariant bases as pointed out by Mansouri et al. \cite{Mansouri}.  In this representation, the
metric tensor of the manifold is block diagonal. In the standard
gauging procedure \cite{Mansouri}, the diagonal form and the basis were given, by
definition, in a sharp contrast with the NLR procedure. To
show clearly this statement,  we give below  the line element
describing the 10-dimensional spacetime manifold, that is  
\begin{eqnarray}
ds^{2}=F_{1}(x,z,\overline{z})W^{2}+F_{2}(x,z,\overline{z})\Sigma ^{2}+%
\overline{F}_{2}(x,z,\overline{z})\overline{\Sigma}^{2},
\end{eqnarray}
where 
\begin{eqnarray}
&&\left( W^{\alpha\overset{\cdot}{\alpha}}\right) ^{2} =\left( 1+\eta
^{2}\right) ^{2}\left[ dx^{\alpha\overset{.}{\alpha}}+N_{\text{ \ }%
\sigma\beta}^{\alpha\overset{.}{\alpha}}dz^{\sigma\beta}+N_{\text{ \ }%
\overset{.}{\sigma}\overset{.}{\gamma}}^{\alpha\overset{.}{\alpha}}d\overset{%
\_}{z}^{\overset{.}{\sigma}\overset{.}{\gamma}}\right] g_{\alpha\overset{%
\cdot}{\alpha}\beta\overset{\cdot}{\beta}}  \notag \\
&&\times\left[ dx^{\beta\overset{.}{\beta}}+N_{\text{ \ }\rho\upsilon}^{\beta%
\overset{.}{\beta}}dz^{\rho\upsilon}+N_{\text{ \ }\overset{.}{\rho}\overset{.%
}{\nu}}^{\beta\overset{.}{\beta}}d\overset{\_}{z}^{\overset{.}{\rho}\overset{%
.}{\nu }}\right] \,, \\
&& \left( \Sigma^{\alpha\beta}\right) ^{2} =\left( 1+\eta^{2}\right)
^{2}\left\{ ie^{-2i\alpha}\left[ -\frac{1}{2}N_{\text{ \ }\rho\overset{.}{%
\alpha}}^{\alpha\beta}dx^{\rho\overset{.}{\alpha}}+dz^{\alpha\beta }\right]
\right\} g_{\left( \alpha\delta\right) \left( \gamma \beta\right) }
\notag \\
&& \times\left\{ ie^{-2i\alpha}\left[ -\frac{1}{2}N_{\text{ \ }\alpha\overset%
{.}{\alpha}}^{\delta\gamma}dx^{\alpha\overset{.}{\alpha}}+dz^{\overset{.}{%
\delta}\overset{\cdot}{\gamma}}\right] \right\}\,, \\
&&\left( \overline{\Sigma}^{\overset{.}{\alpha}\overset{.}{\beta}}\right)
^{2} =\left( 1+\eta^{2}\right) ^{2}\left\{ ie^{2i\alpha}\left[ -\frac{1}{2}%
N_{\text{ \ }\rho\overset{.}{\kappa}}^{\overset{.}{\alpha}\overset{.}{\beta}%
}dx^{\rho\overset{.}{\kappa}}-d\overset{\_}{z}^{\overset{.}{\alpha}\overset{.%
}{\beta}}\right] \right\} g_{\left( \overset{.}{\alpha}\overset{.}{\beta}%
\right) \left( \overset{.}{\delta}\overset{\cdot }{\gamma}\right) } 
\notag \\
&& \times\left\{ ie^{2i\alpha}\left[ -\frac{1}{2}N_{\text{ \ }\alpha\overset{%
.}{\lambda}}^{\overset{.}{\delta}\overset{\cdot}{\gamma}}dx^{\alpha\overset{.%
}{\lambda}}-d\overline{z}^{\overset{.}{\delta}\overset{\cdot}{\gamma}}\right]
\right\}
\end{eqnarray}
with%
\begin{eqnarray}
&& N_{\text{ \ }\sigma\beta}^{\alpha\overset{.}{\alpha}}\equiv\frac {%
2ie^{-2i\alpha}\eta_{\sigma}^{\ \ \overset{.}{\alpha}}\delta_{\beta}^{\alpha
}}{\left( 1+\eta^{2}\right) },\text{ \ }N_{\text{ \ }\overset{.}{\sigma }%
\overset{.}{\gamma}}^{\alpha\overset{.}{\alpha}}\equiv\frac{2ie^{2i\alpha
}\eta_{\ \ \overset{.}{\sigma}}^{\alpha}\delta_{\text{ \ }\overset{.}{\gamma}%
}^{\overset{.}{\alpha}}}{\left( 1+\eta^{2}\right) },  \notag \\
&& \text{ \ }N_{\text{ \ }\rho\overset{.}{\alpha}}^{\alpha\beta}\equiv\frac{%
2ie^{2i\alpha}\eta_{\ \ \overset{.}{\alpha}}^{\beta}\delta_{\rho}^{\alpha}}{%
\left( 1+\eta^{2}\right) },\text{ \ \ }N_{\text{ \ }\rho\overset{.}{\alpha}%
}^{\overset{.}{\sigma}\overset{.}{\gamma}}\equiv\frac{2ie^{-2i\alpha}\eta_{%
\rho}^{\ \ \overset{.}{\sigma}}\delta_{\text{ \ }\overset{.}{\alpha}}^{%
\overset{.}{\gamma}}}{\left( 1+\eta^{2}\right) }\,,
\end{eqnarray}
and 
\begin{eqnarray}
F_{1}(x,z,\overline{z}) & =1-2\chi^{-2}\left( W^{-1}\phi\right) ^{2}\,, \\
F_{2}(x,z,\overline{z}) & =1-2\chi_{2}^{-2}\left( \Sigma^{-1}\alpha\right)
^{2}\,, \\
\overline{F}_{2}(x,z,\overline{z}) & =1-2\chi_{2}^{-2}\left( \overline {%
\Sigma}^{-1}\alpha\right) ^{2}\,.
\end{eqnarray}
In general $g_{\alpha\overset{\cdot}{\alpha}\beta\overset{\cdot}{\beta}}$ $%
g_{\alpha\delta\gamma\beta}$ are proportional to $\epsilon_{\overset{\cdot }{%
\alpha}\overset{\cdot}{\beta}}\epsilon_{\alpha\beta}$ and $\epsilon
_{\alpha\delta}\epsilon_{\gamma\beta}$ ({\it e.g.} in Cartan-Killing metric, the
conjugation operation is assumed in order to be consistent with the scalar
product). The metric is evidently diagonal in $W,\Sigma,\overline{\Sigma}$
and generalize the Nordstrom-Kaluza construction viewed from the coordinate
basis. In general, the simplest basis in the base
manifold is the coordinate one and this is that is  used as standard. For the
bundle viewpoint, it is not convenient to work in a coordinate basis because the
description of isometries in such manifolds can best be carried out in terms
of an invariant basis (giving commutation relations  with the Killing vectors).

It is quite evident tha the field $\eta_{\ \ \overset{.}{\sigma}}^{\alpha}$
plays a central role into $N_{\text{ \ }\overset{.}{\sigma}\overset{.}{\gamma}%
}^{\alpha\overset{.}{\alpha}}$ , the gauge potential of the fiber. In
practical cases, consistently with the algebraic role of $\eta_{\ \ \overset{.%
}{\sigma}}^{\alpha}$, we have the  freedom to select a particular forms for the
gauge potentials $N_{\text{ \ }\overset{.}{\sigma}\overset{.}{\gamma}%
}^{\alpha\overset{.}{\alpha}}$ as the fundamental physical dynamical
variables under consideration. It is also important to consider that, in general, it  is quite obvious that
all the geometrical quantities of interest can be splitted but weakly and
without a complete  reduction. It is worth noticing that supersymmetric cases are more involved in this sense.

\section{The geometrical Lagrangian}

\label{nove} 
Our task is now to define a self-consistent geometrical Lagrangian considering the above results.
As stated, the field $\eta_{\ \ \overset{.}{\sigma}}^{\alpha}$ plays a central role into the gauge potential of the
fiber $N_{\text{ \ }%
B}^{A}$  defined now in the variables $A$, see Sect. \ref{quattro}. This means that  the
geometrical Lagrangian can be simply achieved as  the wedge product of forms. One   can consider 
the standard Einstein-Hilbert Lagrangian, linear in the Ricci scalar $R$, $f(R)$ Lagrangians or quadratic combinations of curvature invariants. For example
\begin{equation}
\sqrt{-G}R;\text{ }\sqrt{-G}R^{\alpha}\left( \alpha\in\mathbb{R}\right) ;%
\text{ }\sqrt{-G}R^{AB}R_{AB};\qquad etc\,.
\end{equation}
Furthermore, Eddington-like Lagrangians can be assumed,  
\begin{equation}
\sqrt{Det(G_{AB}+f\left( R_{AB}\right) )}; \qquad etc\,.
\end{equation}
This  means that any invariant of the Riemann tensor must be into the action:
if it is not the case, there are no dynamical equations for $N_{\text{ \ }%
B}^{A}$ then, for $\eta _{\ \ \overset{.}{\sigma }}^{\alpha }$. In fact,
although the measure $\sqrt{-G}$ contains the simplest geometrical invariant, it  cannot be, from the 
dynamical point of view,  the geometrical Lagrangian. However, it can be used
as a preliminar geometrical object to be analyzed. In the simplest form, it can be
defined as the wedge product of the Cartan forms
\begin{gather*}
\overset{\rho\overset{.}{\rho}=1...4}{\overbrace{\omega_{P}^{\rho\overset{.}{%
\rho}}\wedge....\omega_{P}^{\sigma\overset{.}{\sigma}}}}\wedge \overset{%
\alpha\beta\left( \overset{.}{\alpha}\overset{.}{\beta}\right) =5...10}{%
\overbrace{\omega_{Z}^{\alpha\beta}\wedge....\omega_{Z}^{\delta
\gamma}\wedge\omega_{\overline{Z}}^{\overset{.}{\alpha}\overset{.}{\beta}%
}\wedge....\omega_{\overline{Z}}^{\overset{.}{\delta}\overset{.}{\gamma}}}}=
\\
=\left( me^{\phi}\frac{1+\eta^{2}}{1-\eta^{2}}\right) ^{10}\underset{a}{%
\underbrace{\frac{1-\frac{\xi^{2}}{2}}{1+\frac{\xi^{2}}{2}}}}\ \ \ \underset{%
b}{\underbrace{\frac{1-\frac{\overline{\xi}^{2}}{2}}{1+\frac{\overline{\xi}%
^{2}}{2}}}}\underset{c}{\underbrace{\frac{1-\lambda ^{2}/2}{1+\lambda^{2}/2}}%
}
\end{gather*}
This wedge product can be written as a function of $\eta$ (and the scalar
and pseudo-scalar $\phi$ and $\alpha$ respectively). To this end, we 
write the factors $\ b$ and $c$ of the above equation explicitly (for$\ a$
the dependence is clear) defining firstly%
\begin{eqnarray}
& \left[ \left( W^{-1}\right) _{\alpha\overset{.}{\alpha}}\phi\right] \left[
\left( W^{-1}\right) ^{\alpha\overset{.}{\alpha}}\phi\right] \equiv\left(
W^{-1}\phi\right) ^{2}\,,  \notag \\
\notag \\
& \left[ \left( \Sigma^{-1}\right) _{\alpha\beta}\alpha\right] \left[ \left(
\Sigma^{-1}\right) ^{\alpha\beta}\alpha\right] \equiv\left(
\Sigma^{-1}\alpha\right) ^{2}\,,
\end{eqnarray}
then 
\begin{eqnarray}
c & = \sqrt{1-2\chi^{-2}\left( W^{-1}\phi\right) ^{2}}\,,  \notag \\
\notag \\
a & = \sqrt{1-2\chi_{2}^{-2}\left( \Sigma^{-1}\alpha\right) ^{2}}\,,  \notag
\\
\notag \\
b & =\sqrt{1-2 \c hi_{2}^{-2}\left( \overline{\Sigma}^{-1}\alpha\right) \,,
^{2}}
\end{eqnarray}
for instance, and due the simple assumption $\xi _{\alpha \beta }=$ $\xi
_{\alpha \beta }\left( \alpha \right) $ and $\lambda _{\alpha \overset{.}{%
\alpha }}=\lambda _{\alpha \overset{.}{\alpha }}\left( \phi \right) $, the
geometrical Lagrangian takes the  form%
\begin{equation*}
L_{G}=\left( me^{\phi }\frac{1+\eta ^{2}}{1-\eta ^{2}}\right) ^{10}\sqrt{%
\left[ 1-2\chi _{2}^{-2}\left( \Sigma ^{-1}\alpha \right) ^{2}\right] \left[
1-2\chi _{2}^{-2}\left( \overline{\Sigma }^{-1}\alpha \right) ^{2}\right] %
\left[ 1-2\chi _{1}^{-2}\left( W^{-1}\phi \right) ^{2}\right] }\,,
\end{equation*}%
where the dynamics of the axion, dilaton and other fields are governed by a
Born-Infeld-like action. It is evident the interplay of  gravity and
fields dynamics. The case is similar to the  picture given by the
Weyl theory where the scalar field is associated to some part of the
connection. Finally, we can state that any suitable geometrical Lagrangian, constructed by   suitable combination of curvature invariants in such an approach, can contain, in principle,   the dynamics of  gravitational field and  all the information about  matter fields
and their interactions.

\section{Conclusions and Perspectives}

\label{dieci} 

In this paper a nonlinearly realized representation of the local
conformal-affine group has been determined. Gravity and spin are the results
of such a realization. It has been found that the nonlinear Lorentz
transformation law contains contributions from the linear Lorentz parameters
as well as conformal and shear contributions via the nonlinear $4$-boosts
and symmetric $GL(4)$ parameters. We have identified the pullback of the
nonlinear translational connection coefficient to a manifold $M$ as a space-time
coframe. In this way, the frame fields of the theory are obtained from the
(nonlinear) gauge prescription. The mixed index coframe components (tetrads)
are used to convert from the Lie algebra indices to space-time indices. In this picture, the
space-time metric is a secondary object constructed (induced!) from the
constant $H$ group metric and the tetrads.

The problem of consistency for the determination of   the form of geometrical
Lagrangian  (that is one of the main drawbacks of the nonlinear
realization approach) is here avoided due to the exact identification of the
structure of the fiber (gauge potentials). The Lagrangian must generate the
corresponding dynamical equations of motion for these gauge potentials.This
fact have to be pointed out because such an  identification, in our knowledge, has never been 
 achieved in a clear form. 

As a concluding remark, we can say that gravity (and in general any gauge
field) can be derived as the nonlinear realization of a local
conformal-affine symmetry group and then gravity can be considered an
interaction induced from invariance properties. In a forthcoming paper, we will study in detail the geometrical Lagrangian and the related field equations in order to point out the physical properties of such an approach.

\section*{Acknowledgements}

SC and MDL acknowledge the support of INFN Sez. di Napoli (Iniziative
Specifiche TEONGRAV and QGSKY). D. J. C-L. thanks JINR-BLTP (Russia) for hospitality
and financial support.

\end{document}